\documentclass[aps,article,notitlepage,superscriptaddress,prd,floatfix,showpacs,preprintnumbers,amsmath,amssymb,letterpaper]{revtex4-1}
\newcommand{\beq}{\begin{equation}}
\newcommand{\eeq}{\end{equation}}
\newcommand{\beqn}{\begin{eqnarray}}
\newcommand{\eeqn}{\end{eqnarray}}

\usepackage{graphicx}
\usepackage{epsf}
\usepackage{graphics,epsfig,placeins,subfigure,wrapfig,amssymb,pifont}
\usepackage{hyperref}
\hypersetup{colorlinks=true, urlcolor=blue}

\begin{document}
\title{An efficient spectral interpolation routine for the {\tt TwoPunctures} code}

\author{Vasileios Paschalidis}
\author{Zachariah B. Etienne}
\author{Roman Gold}
\author{Stuart~L.~Shapiro}
\altaffiliation{Also at Department of Astronomy and NCSA, University of
  Illinois at Urbana-Champaign, Urbana, IL 61801}
\affiliation{Department of Physics, University of Illinois at
  Urbana-Champaign, Urbana, IL 61801}

\begin{abstract}

{\tt TwoPunctures} is perhaps the most widely-adopted code for
generating binary black hole ``puncture'' initial data and
interpolating these (spectral) data onto evolution grids. In typical
usage, the bulk of this code's run time is spent in its spectral
interpolation routine. We announce a new publicly-available spectral
interpolation routine that improves the performance of the original
interpolation routine by a factor of $\sim$100, yielding results
consistent with the original spectral interpolation routine to
roundoff precision. This note serves as a guide for installing this
routine both in the original standalone {\tt TwoPunctures} code and
the Einstein Toolkit supported version of this code.

\end{abstract}


\maketitle

\section{Introduction}

One of the pressing goals of numerical relativity is to calculate
accurate gravitational waveforms from plausible astrophysical sources
to help generate templates that will be used by ground based
gravitational wave observatories such as LIGO \cite{LIGO1,LIGO2},
VIRGO \cite{VIRGO1,VIRGO2}, TAMA \cite{TAMA1,TAMA2},GEO \cite{GEO},
KAGRA \cite{LCGT}, and by proposed space-based interferometers such as
eLISA/NGO \cite{eLISA} and DECIGO \cite{DECIGO}. This task is far from
trivial and the very first step in accomplishing it is the generation
of initial data that satisfy the Einstein constraints \cite{BSBook}.

To date there are multiple codes that solve the constraints of
Einstein's theory of general relativity (see
e.g. \cite{Pfeiffer,LORENE,Paschalidis:2010dh,Paschalidis:2011ez,EAST,TwoPunctures}
and references therein).  As the inspiral and merger of compact
binaries such as binary black holes (BHBH), binary neutron stars and
binary black hole--neutron stars, furnish some of the most
promising astrophysical scenarios (both in terms of signal strength
and event rates) for the generation of detectable gravitational waves,
the solutions obtained using these initial data solvers are mainly
focused on compact binary systems.

Most of the recent focus in gravitational wave template generation has
been on binary black hole systems (see e.g. \cite{Ajith,NINJA2}) and for this
reason one of the most popular codes for generating BHBH initial data
is the {\tt TwoPunctures} code \cite{TwoPunctures}, which has also been
adopted by the publicly available Einstein Toolkit \cite{ET}. The
popularity of this code stems from the fact that it is remarkably
user-friendly, spectrally accurate and efficient in solving the
Einstein constraints for BHBHs when both BHs are represented as
punctures. Once the initial value problem has been solved, the initial data
have to be mapped onto the dynamical evolution grids via
interpolation. The {\tt TwoPunctures} code offers two interpolation
routines: i) a second-order polynomial interpolation routine, and ii)
a spectral interpolation routine.  The former is very fast but we have
found empirically that it is not well-suited for dynamical
evolutions. For this reason, it is the latter that is most widely used
by us and most numerical relativity groups.

In this brief note we provide a spectral interpolation routine for the
{\tt TwoPunctures} code (and installation instructions both for the
Einstein Toolkit version of the code, i.e., the {\tt TwoPunctures}
thorn, and its standalone version) that is $\sim100$ times faster than
the original spectral interpolation routine of the {\tt TwoPunctures}
code. This new routine saves many hours of computation, especially
when high spectral resolutions are used. By no means do we claim that
we have optimized the process, and one may find other ways of
optimizing the performance of the code in general (see
below). However, we find that the acceleration attained by our routine
is sufficiently satisfactory, and we hope that by making this routine
publicly available, the many numerical relativity groups that use
TwoPunctures will benefit from this faster spectral interpolation
routine.

\section{A faster spectral interpolation routine}

Our basic modification of the {\tt TwoPunctures} code was stimulated
by the observation that each time the original {\tt TwoPunctures} code
spectral interpolation routine is called it computes the spectral
interpolation coefficients, given the values of the function at the
collocation points, and then uses the spectral expansion to
interpolate to any one Cartesian grid point. Typical high resolution
evolution grids in finite difference codes employ 8-9 levels of
refinement with resolutions of $M/40$ or higher, where $M$ is the BHBH
ADM mass. This amounts to a grid of about $10^5-10^6$ zones for which
interpolations must be performed. Calculating the spectral
coefficients from the values at the collocation points is an expensive
operation, which is exacerbated by computing them at every
interpolated point.
In our initial value calculations we use the {\tt TwoPunctures} code
with $50\times 70^2$ basis functions, thus the original {\tt
  TwoPunctures} code spectral interpolation routine calculates $\sim
2.5\times 10^5$ spectral coefficients each time it is called, i.e.,
$\sim 10^{6}$ times. This means that the spectral coefficients are
recomputed $\sim 10^{11}$ times in this process.

\begin{figure*}
\centering
\includegraphics[width=0.49\textwidth]{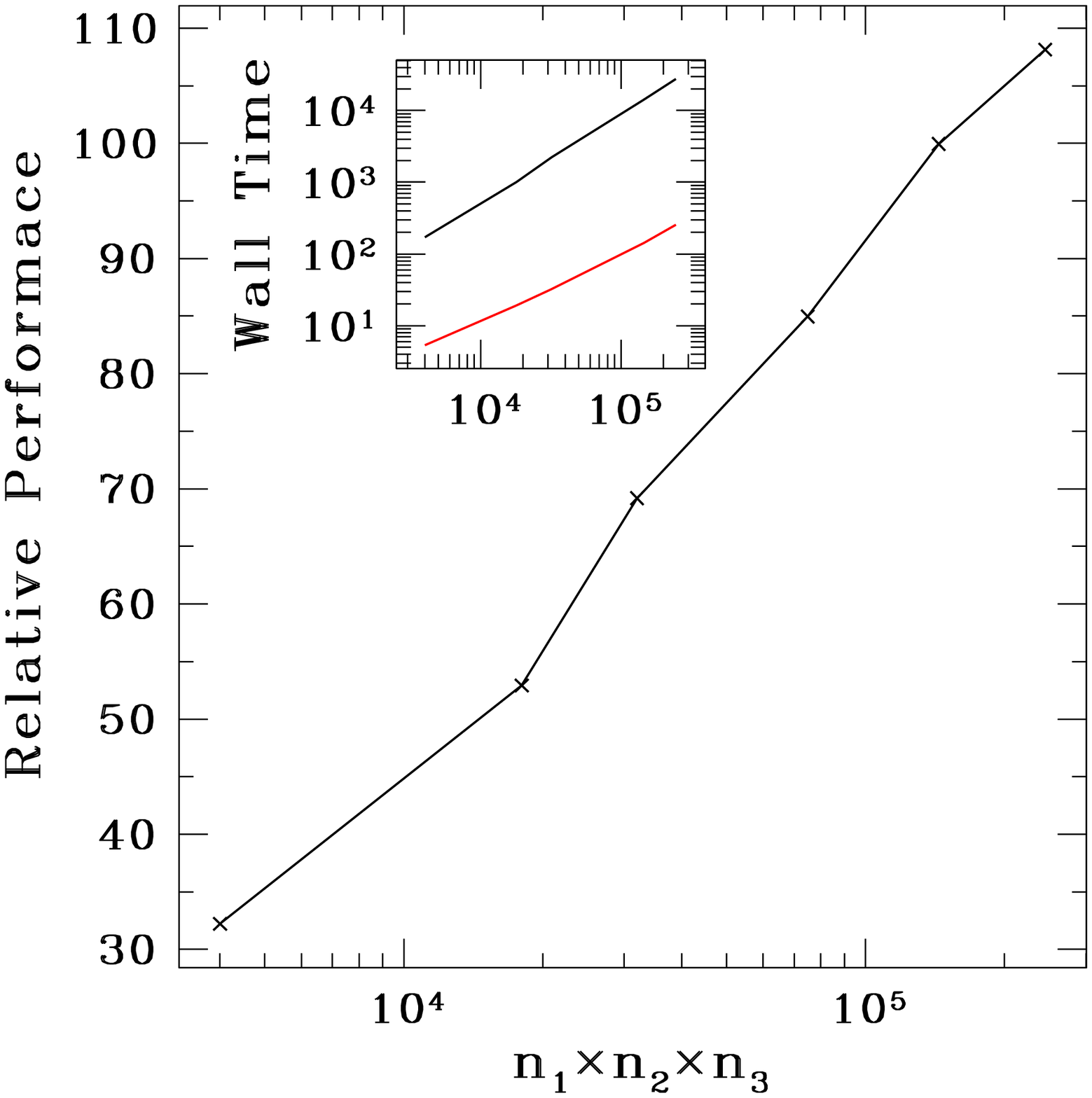}
\includegraphics[width=0.49\textwidth]{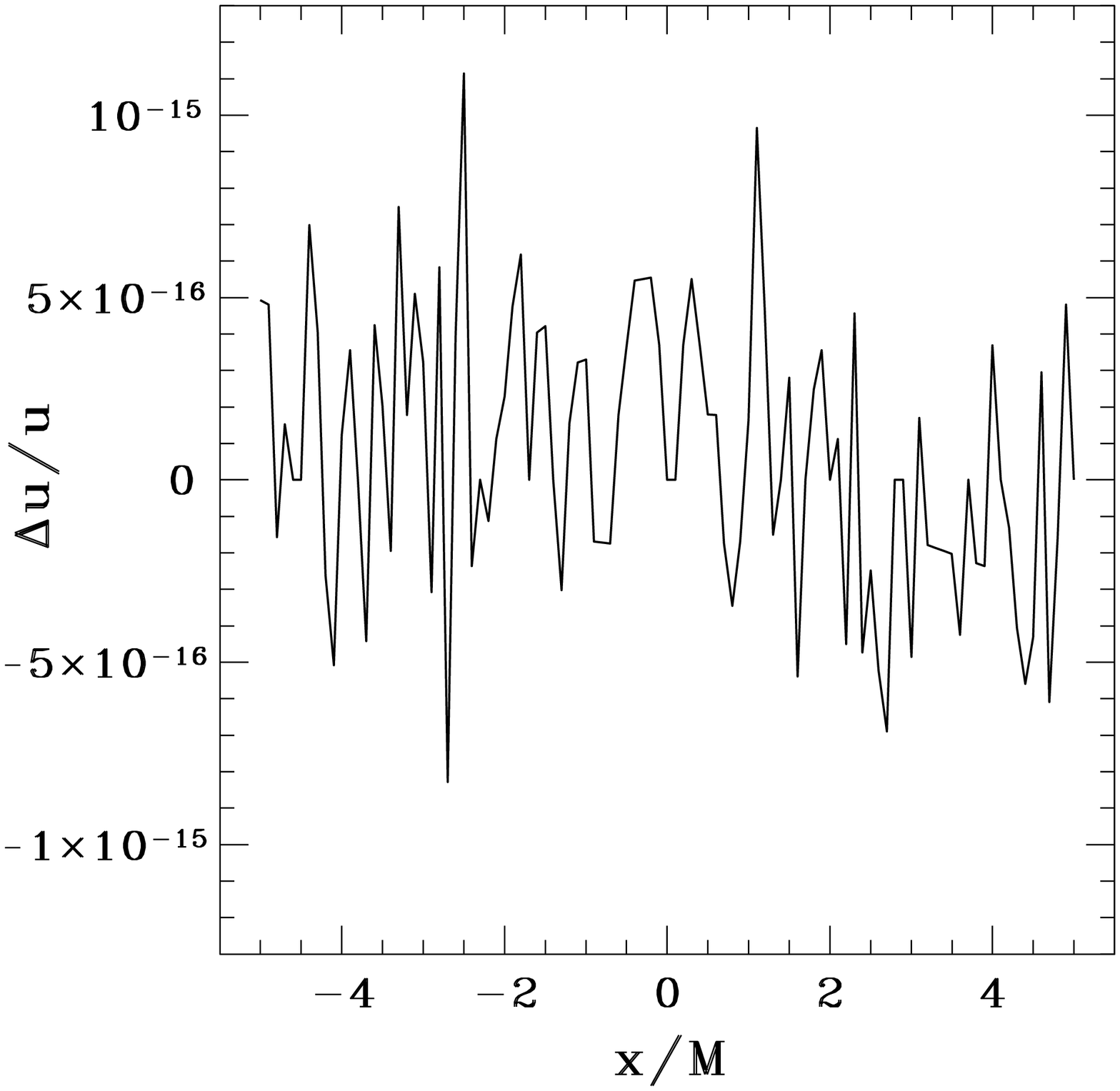}
\caption{Left:Performance of the original {\tt TwoPunctures} code
  interpolation routine normalized by the performance of our new spectral
  interpolation routine versus the total number of basis functions (or
  collocation points). Performance is measured in seconds of wall
  time. The inset shows the required wall time in seconds. The black
  curve corresponds to the original {\tt TwoPunctures} code
  interpolation routine and the red curve to our spectral
  interpolation routine. The performance test is based on timing the
  interpolation operation for a grid consisting of $10^5$ zones, after
  the code has solved for a particular BHBH configuration.  The
  spectral resolutions used were $(n_1, n_2, n_3)=$ $(10,20,20)$,
  $(20, 30, 30)$, $(20, 40, 40)$, $(30, 50, 50)$, $(40, 60, 60)$,
  $(50, 70, 70)$.  Right: Fractional difference between the
  interpolated values along the BHBH binary axis (the punctures lie at
  $x/M=\pm 2.17$) using the original {\tt TwoPuncture} code
  interpolation routine and our new spectral interpolation
  routine. The results agree to machine precision.  All runs were
  performed on a system with an Intel Core 2 Duo 6300 processor.  The
  code was compiled with the Intel 11.1 C++ compiler with -O3
  optimizations.  Similar relative performance is found on other
  systems and with other compiler optimizations.  }
\label{performance}
\end{figure*}

The total cost of this operation can be reduced significantly, if the
spectral coefficients computation is a one-time operation. So, our new
interpolation routine uses the Chebyshev and Fourier basis routines in
the {\tt TwoPunctures} code to compute the spectral coefficients once
and for all, and then store them in an array. Another routine we
developed takes the stored spectral coefficients as input and
performs the spectral interpolation. To compute the required sums we
use the partial summation method \cite{Boyd}. Our new routine still
computes the values of the bases functions at the collocation points
in order to perform the interpolation every time it is called, hence
further optimization can take place by storing the values of the bases
functions at the collocation points as well. However, we did not do so
because the new interpolation routine performs sufficiently fast.

In the left panel of Fig.~\ref{performance} we show the performance of
the original {\tt TwoPunctures} spectral interpolation routine
normalized by the performance of our routine against the total number
of collocation points for a test case where we do $10^5$
interpolations after solving for an initial BHBH configuration. It is
clear that our routine accelerates the interpolation procedure. The
higher the spectral resolution the greater the gain, and for
resolutions of $50\times 70^2$ we get more than a factor of 100
speed-up: our routine takes $\sim 200$s, while the original routine
requires $\sim 28000$s to finish. We have also checked that the
results with the two routines agree to machine precision as shown in
the right panel of Fig.~\ref{performance}.

\section{Instructions for installing the new interpolation routine}

Here we provide instructions for installing the new interpolation
routine both in the {\tt TwoPunctures} thorn of the Einstein Toolkit,
and the standalone version of the {\tt TwoPunctures} code. The
routines for the Einstein Toolkit version of the {\tt TwoPunctures}
code can be downloaded from
\\ \url{http://webusers.physics.illinois.edu/~vpaschal/TwoPuncturesET/}.
The routines for the standalone \\ {\tt TwoPunctures}
code can be downloaded from
\\ \url{http://webusers.physics.illinois.edu/~vpaschal/TwoPunctures_Standalone/}.

\subsection{{\tt TwoPunctures} Einstein Toolkit}
\label{TwoPuncturesET}

We provide the routines 

\begin{verbatim}
void SpecCoef(int n1, int n2, int n3, int ivar, CCTK_REAL *v, CCTK_REAL *cf)

CCTK_REAL PunctIntPolAtArbitPositionFast(int ivar, int nvar, int n1,
                                         int n2, int n3, derivs v, 
                                         CCTK_REAL x, CCTK_REAL y, CCTK_REAL z)

CCTK_REAL PunctEvalAtArbitPositionFast(CCTK_REAL *v, int ivar, CCTK_REAL A, 
                                       CCTK_REAL B, CCTK_REAL phi, int nvar, 
                                       int n1, int n2, int n3)
\end{verbatim}

The first computes the spectral expansion coefficients ({\tt cf}) of a
variable ({\tt v}) given the values of variable {\tt v} at the
collocation points, where {\tt n1, n2, n3}, are the number of basis
functions in the $A,\ B,\ \phi$ coordinates of the {\tt TwoPunctures}
code. The second routine applies spectral interpolation on the
variable {\tt v} to the desired point with Cartesian coordinates
$x,y,z$, using the third routine which interpolates {\tt v} using the
corresponding {\tt TwoPunctures} coordinates $A, B, \phi$. The
following are the required steps to implement our new interpolation
routine.

\begin{enumerate}

\item For convenience we recommend that these routines be added in the file
FuncAndJacobian.c of the {\tt TwoPunctures} thorn.

\item These routines must also be declared in the TwoPunctures.h header file.


\item In the file TwoPuncture.c the following declaration statement must be added:
\begin{verbatim}
 static derivs cf_v;
\end{verbatim}

\item In the file TwoPuncture.c the following memory allocation call
  for storage of the spectral coefficients must be added:
\begin{verbatim}
allocate_derivs (&cf_v, ntotal);
\end{verbatim}

\item Following memory allocation, the new variables should be initialized (e.g.):

\begin{verbatim}
    for (int j = 0; j < ntotal; j++)
    {
      cf_v.d0[j] = 0.0;  cf_v.d1[j] = 0.0;
      cf_v.d2[j] = 0.0;  cf_v.d3[j] = 0.0;
      cf_v.d11[j] = 0.0; cf_v.d12[j] = 0.0;
      cf_v.d13[j] = 0.0; cf_v.d22[j] = 0.0;
      cf_v.d23[j] = 0.0; cf_v.d33[j] = 0.0;
    }
\end{verbatim}
 
Note that there is a redundancy, as {\tt cf\_v.d0} is the only required variable, but as the memory footprint of the {\tt cf\_v} derivs struct 
is small this redundancy is of minor significance.

\item In the file TwoPuncture.c following the function call  

\begin{verbatim}
F_of_v (cctkGH, nvar, n1, n2, n3, v, F, u);
\end{verbatim}
which follows the loop checking for convergence of the puncture masses,
the call to the calculation of the spectral coefficients 
should be made

\begin{verbatim}
SpecCoef(n1, n2, n3, 0, v.d0, cf_v.d0);
\end{verbatim}

\item All calls to {\tt PunctIntPolAtArbitPosition} should be replaced 
by corresponding calls to \\ {\tt PunctIntPolAtArbitPositionFast}, but 
the most important change is to replace the call 

\begin{verbatim}
U = PunctIntPolAtArbitPosition(0, nvar, n1, n2, n3, v, x1, y1, z1);
\end{verbatim}

by the call

\begin{verbatim}
U = PunctIntPolAtArbitPositionFast(0, nvar, n1, n2, n3, cf_v, x1, y1, z1);
\end{verbatim}

\item Memory for the spectral coefficients should be freed at the end of the TwoPuncture.c file

\begin{verbatim}
free_derivs (&cf_v, ntotal);
\end{verbatim}

\end{enumerate}

\subsection{Standalone {\tt TwoPunctures} code}

For the standalone {\tt TwoPunctures} code we provide the following routines:

\begin{verbatim}

void SpecCoef(parameters par, int ivar, double *v, double *cf)
double Spec_IntPolFast (parameters par, int ivar, double *v, double x, double y, double z)
double Spec_IntPolABphiFast (parameters par, double *v, int ivar, double A, double B, double phi)

\end{verbatim}

These routines do precisely the same calculations as the routines {\tt
  SpecCoef, PunctIntPolAtArbitPositionFast,
  PunctEvalAtArbitPositionFast} in the {\tt TwoPuncturesET} version of
the code.

The steps to implement our new interpolation routine in the standalone version 
are similar to the ones we outlined in the previous section \ref{TwoPuncturesET}

\begin{enumerate}

\item These routines can be added to the file FuncAndJacobian.C of the {\tt TwoPunctures} code.

\item These routines must be declared in the TwoPunctures.h header file.


\item In the file TwoPunctures.C the following declaration statement must be added:
\begin{verbatim}
 derivs cf_v;
\end{verbatim}

\item In the file TwoPunctures.C memory should be allocated for storage of the spectral coefficients:
\begin{verbatim}
allocate_derivs (&cf_v, ntotal);
\end{verbatim}

\item In the file TwoPunctures.C following the call  
\begin{verbatim}
Newton(par, v);
\end{verbatim}

the calculation of the spectral coefficients should be performed:
\begin{verbatim}
SpecCoef(n1, n2, n3, 0, v.d0, cf_v.d0);
\end{verbatim}

\item Calls to
\begin{verbatim}
Spec_IntPol(par, 0, v.d0, x, y, z);
\end{verbatim}
should be replaced by calls to
\begin{verbatim}
Spec_IntPolFast(par, 0, cf_v.d0, x, y, z);
\end{verbatim}

\item Memory for the spectral coefficients should be freed at the end of the TwoPunctures.C file

\begin{verbatim}
free_derivs (&cf_v, ntotal);
\end{verbatim}

\end{enumerate}

\acknowledgments

This work was supported in part by NSF Grants PHY-0963136 as well as
NASA Grant NNX11AE11G to the University of Illinois at
Urbana-Champaign. Z. Etienne gratefully acknowledges support from NSF
Astronomy and Astrophysics Postodoctoral Fellowship AST-1002667.

\bibliography{paper}

\begin{thebibliography}{10}%
\makeatletter
\providecommand \@ifxundefined [1]{%
 \ifx #1\undefined \expandafter \@firstoftwo
 \else \expandafter \@secondoftwo
\fi
}%
\providecommand \@ifnum [1]{%
 \ifnum #1\expandafter \@firstoftwo
 \else \expandafter \@secondoftwo
\fi
}%
\providecommand \enquote [1]{``#1''}%
\providecommand \bibnamefont  [1]{#1}%
\providecommand \bibfnamefont [1]{#1}%
\providecommand \citenamefont [1]{#1}%
\providecommand\href[0]{\@sanitize\@href}%
\providecommand\@href[1]{\endgroup\@@startlink{#1}\endgroup\@@href}%
\providecommand\@@href[1]{#1\@@endlink}%
\providecommand \@sanitize [0]{\begingroup\catcode`\&12\catcode`\#12\relax}%
\@ifxundefined \pdfoutput {\@firstoftwo}{%
 \@ifnum{\z@=\pdfoutput}{\@firstoftwo}{\@secondoftwo}%
}{%
 \providecommand\@@startlink[1]{\leavevmode\special{html:<a href="#1">}}%
 \providecommand\@@endlink[0]{\special{html:</a>}}%
}{%
 \providecommand\@@startlink[1]{%
  \leavevmode
  \pdfstartlink
   attr{/Border[0 0 1 ]/H/I/C[0 1 1]}%
   user{/Subtype/Link/A<</Type/Action/S/URI/URI(#1)>>}%
  \relax
 }%
 \providecommand\@@endlink[0]{\pdfendlink}%
}%
\providecommand \url  [0]{\begingroup\@sanitize \@url }%
\providecommand \@url [1]{\endgroup\@href {#1}{\urlprefix}}%
\providecommand \urlprefix [0]{URL }%
\providecommand \Eprint[0]{\href }%
\@ifxundefined \urlstyle {%
  \providecommand \doi [1]{doi:\discretionary{}{}{}#1}%
}{%
  \providecommand \doi [0]{doi:\discretionary{}{}{}\begingroup
  \urlstyle{rm}\Url }%
}%
\providecommand \doibase [0]{http://dx.doi.org/}%
\providecommand \Doi[1]{\href{\doibase#1}}%
\providecommand \bibAnnote [3]{%
  \BibitemShut{#1}%
  \begin{quotation}\noindent
    \textsc{Key:}\ #2\\\textsc{Annotation:}\ #3%
  \end{quotation}%
}%
\providecommand \bibAnnoteFile [2]{%
  \IfFileExists{#2}{\bibAnnote {#1} {#2} {\input{#2}}}{}%
}%
\providecommand \typeout [0]{\immediate \write \m@ne }%
\providecommand \selectlanguage [0]{\@gobble}%
\providecommand \bibinfo [0]{\@secondoftwo}%
\providecommand \bibfield [0]{\@secondoftwo}%
\providecommand \translation [1]{[#1]}%
\providecommand \BibitemOpen[0]{}%
\providecommand \bibitemStop [0]{}%
\providecommand \bibitemNoStop [0]{.\EOS\space}%
\providecommand \EOS [0]{\spacefactor3000\relax}%
\providecommand \BibitemShut [1]{\csname bibitem#1\endcsname}%
\bibitem{LIGO1}%
  \BibitemOpen
  \bibfield{author}{%
  \bibinfo {author} {\bibfnamefont{B.}~\bibnamefont{{Abbott}}}\ and\ \bibinfo
  {author} {\bibnamefont{{the LIGO Scientific Collaboration}}},\ }%
  \bibfield{journal}{%
  \Doi{10.1103/PhysRevD.77.062002}{\bibinfo {journal} {\prd}}\ }%
  \textbf{\bibinfo {volume} {77}},\ \bibinfo {pages} {062002} (\bibinfo {month}
  {Mar.}\ \bibinfo {year} {2008})%
  \bibAnnoteFile{NoStop}{LIGO1}%
\bibitem{LIGO2}%
  \BibitemOpen
  \bibfield{author}{%
  \bibinfo {author} {\bibfnamefont{D.~A.}\ \bibnamefont{{Brown}}}, \bibinfo
  {author} {\bibfnamefont{S.}~\bibnamefont{{Babak}}}, \bibinfo {author}
  {\bibfnamefont{P.~R.}\ \bibnamefont{{Brady}}}, \bibinfo {author}
  {\bibfnamefont{N.}~\bibnamefont{{Christensen}}}, \bibinfo {author}
  {\bibfnamefont{T.}~\bibnamefont{{Cokelaer}}}, \bibinfo {author}
  {\bibfnamefont{J.~D.~E.}\ \bibnamefont{{Creighton}}}, \bibinfo {author}
  {\bibfnamefont{S.}~\bibnamefont{{Fairhurst}}}, \bibinfo {author}
  {\bibfnamefont{G.}~\bibnamefont{{Gonzalez}}}, \bibinfo {author}
  {\bibfnamefont{E.}~\bibnamefont{{Messaritaki}}}, \bibinfo {author}
  {\bibfnamefont{B.~S.}\ \bibnamefont{{Sathyaprakash}}}, \bibinfo {author}
  {\bibfnamefont{P.}~\bibnamefont{{Shawhan}}},\ and\ \bibinfo {author}
  {\bibfnamefont{N.}~\bibnamefont{{Zotov}}},\ }%
  \bibfield{journal}{%
  \bibinfo {journal} {Class.~Quant.~Grav.}\ }%
  \textbf{\bibinfo {volume} {21}},\ \bibinfo {pages} {S1625} (\bibinfo {month}
  {Oct.}\ \bibinfo {year} {2004})%
  \bibAnnoteFile{NoStop}{LIGO2}%
\bibitem{VIRGO1}%
  \BibitemOpen
  \bibfield{author}{%
  \bibinfo {author} {\bibfnamefont{F.}~\bibnamefont{{Acernese}}}\ and\ \bibinfo
  {author} {\bibnamefont{{the VIRGO Collaboration}}},\ }%
  \bibfield{journal}{%
  \Doi{10.1088/0264-9381/23/19/S01}{\bibinfo {journal} {Class.~Quant.~Grav.}}\
  }%
  \textbf{\bibinfo {volume} {23}},\ \bibinfo {pages} {S635} (\bibinfo {month}
  {Oct.}\ \bibinfo {year} {2006})%
  \bibAnnoteFile{NoStop}{VIRGO1}%
\bibitem{VIRGO2}%
  \BibitemOpen
  \bibfield{author}{%
  \bibinfo {author} {\bibfnamefont{F.}~\bibnamefont{{Beauville}}}\ and\
  \bibinfo {author} {\bibnamefont{{the LIGO-VIRGO Working Group}}},\ }%
  \bibfield{journal}{%
  \Doi{10.1088/0264-9381/25/4/045001}{\bibinfo {journal} {Classical and Quantum
  Gravity}}\ }%
  \textbf{\bibinfo {volume} {25}},\ \bibinfo {pages} {045001} (\bibinfo {month}
  {Feb.}\ \bibinfo {year} {2008})%
  \bibAnnoteFile{NoStop}{VIRGO2}%
\bibitem{TAMA1}%
  \BibitemOpen
  \bibfield{author}{%
  \bibinfo {author} {\bibfnamefont{M.}~\bibnamefont{{Ando}}}\ and\ \bibinfo
  {author} {\bibnamefont{{the TAMA collaboration}}},\ }%
  \bibfield{journal}{%
  \bibinfo {journal} {Class. Quant. Grav.}\ }%
  \textbf{\bibinfo {volume} {19}},\ \bibinfo {pages} {1409} (\bibinfo {month}
  {Apr.}\ \bibinfo {year} {2002})%
  \bibAnnoteFile{NoStop}{TAMA1}%
\bibitem{TAMA2}%
  \BibitemOpen
  \bibfield{author}{%
  \bibinfo {author} {\bibfnamefont{D.}~\bibnamefont{{Tatsumi}}}\ and\ \bibinfo
  {author} {\bibnamefont{{the TAMA collaboration}}},\ }%
  \bibfield{journal}{%
  \Doi{10.1088/0264-9381/24/19/S03}{\bibinfo {journal} {Classical and Quantum
  Gravity}}\ }%
  \textbf{\bibinfo {volume} {24}},\ \bibinfo {pages} {S399} (\bibinfo {month}
  {Oct.}\ \bibinfo {year} {2007})%
  \bibAnnoteFile{NoStop}{TAMA2}%
\bibitem{GEO}%
  \BibitemOpen
  \bibfield{author}{%
  \bibinfo {author} {\bibfnamefont{H.}~\bibnamefont{{L{\"u}ck}}}\ and\ \bibinfo
  {author} {\bibnamefont{{the GEO600 collaboration}}},\ }%
  \bibfield{journal}{%
  \Doi{10.1088/0264-9381/23/8/S10}{\bibinfo {journal} {Class.~Quant.~Grav.}}\
  }%
  \textbf{\bibinfo {volume} {23}},\ \bibinfo {pages} {S71} (\bibinfo {month}
  {Apr.}\ \bibinfo {year} {2006})%
  \bibAnnoteFile{NoStop}{GEO}%
\bibitem{LCGT}%
  \BibitemOpen
  \bibfield{author}{%
  \bibinfo {author} {\bibfnamefont{K.}~\bibnamefont{{Kuroda}}}\ and\ \bibinfo
  {author} {\bibnamefont{{LCGT Collaboration}}},\ }%
  \bibfield{journal}{%
  \Doi{10.1088/0264-9381/27/8/084004}{\bibinfo {journal} {Classical and Quantum
  Gravity}}\ }%
  \textbf{\bibinfo {volume} {27}},\ \bibinfo {pages} {084004} (\bibinfo {month}
  {Apr.}\ \bibinfo {year} {2010})%
  \bibAnnoteFile{NoStop}{LCGT}%
\bibitem{eLISA}%
  \BibitemOpen
  \bibfield{author}{%
  \bibinfo {author} {\bibfnamefont{P.}~\bibnamefont{Amaro-Seoane}}, \bibinfo
  {author} {\bibfnamefont{S.}~\bibnamefont{Aoudia}}, \bibinfo {author}
  {\bibfnamefont{S.}~\bibnamefont{Babak}}, \bibinfo {author}
  {\bibfnamefont{P.}~\bibnamefont{Binetruy}}, \bibinfo {author}
  {\bibfnamefont{E.}~\bibnamefont{Berti}}, \emph{et~al.}}%
   (\bibinfo {year} {2012}),\
  \Eprint{http://arxiv.org/abs/1201.3621}{arXiv:1201.3621 [astro-ph.CO]}%
  \bibAnnoteFile{NoStop}{eLISA}%
\bibitem{DECIGO}%
  \BibitemOpen
  \bibfield{author}{%
  \bibinfo {author} {\bibfnamefont{S.}~\bibnamefont{{Kawamura}}}\ and\ \bibinfo
  {author} {\bibnamefont{{the DECIGO collaboration}}},\ }%
  \bibfield{journal}{%
  \Doi{10.1088/0264-9381/23/8/S17}{\bibinfo {journal} {Class.~Quant.~Grav.}}\
  }%
  \textbf{\bibinfo {volume} {23}},\ \bibinfo {pages} {S125} (\bibinfo {month}
  {Apr.}\ \bibinfo {year} {2006})%
  \bibAnnoteFile{NoStop}{DECIGO}%
\bibitem{BSBook}%
  \BibitemOpen
  \bibfield{author}{%
  \bibinfo {author} {\bibfnamefont{T.~W.~L.}\ \bibnamefont{Baumgarte}}\ and\
  \bibinfo {author} {\bibfnamefont{S.~L.}\ \bibnamefont{Shapiro}},\ }%
  \emph{\bibinfo {title} {Numerical Relativity}}\ (\bibinfo {publisher}
  {Cambridge University Press},\ \bibinfo {year} {2010})%
  \bibAnnoteFile{NoStop}{BSBook}%
\bibitem{Pfeiffer}%
  \BibitemOpen
  \bibfield{author}{%
  \bibinfo {author} {\bibfnamefont{H.~P.}\ \bibnamefont{Pfeiffer}}, \bibinfo
  {author} {\bibfnamefont{L.~E.}\ \bibnamefont{Kidder}}, \bibinfo {author}
  {\bibfnamefont{M.~A.}\ \bibnamefont{Scheel}},\ and\ \bibinfo {author}
  {\bibfnamefont{S.~A.}\ \bibnamefont{Teukolsky}},\ }%
  \bibfield{journal}{%
  \Doi{10.1016/S0010-4655(02)00847-0}{\bibinfo {journal}
  {Comput.Phys.Commun.}}\ }%
  \textbf{\bibinfo {volume} {152}},\ \bibinfo {pages} {253} (\bibinfo {year}
  {2003}),\ \Eprint{http://arxiv.org/abs/gr-qc/0202096}{arXiv:gr-qc/0202096
  [gr-qc]}%
  \bibAnnoteFile{NoStop}{Pfeiffer}%
\bibitem{LORENE}%
  \BibitemOpen
  \bibinfo {note} {{\tt http://www.lorene.obspm.fr/}}%
  \bibAnnoteFile{NoStop}{LORENE}%
\bibitem{Paschalidis:2010dh}%
  \BibitemOpen
  \bibfield{author}{%
  \bibinfo {author} {\bibfnamefont{V.}~\bibnamefont{Paschalidis}}, \bibinfo
  {author} {\bibfnamefont{Z.}~\bibnamefont{Etienne}}, \bibinfo {author}
  {\bibfnamefont{Y.~T.}\ \bibnamefont{Liu}},\ and\ \bibinfo {author}
  {\bibfnamefont{S.~L.}\ \bibnamefont{Shapiro}},\ }%
  \bibfield{journal}{%
  \Doi{10.1103/PhysRevD.83.064002}{\bibinfo {journal} {Phys.Rev.}}\ }%
  \textbf{\bibinfo {volume} {D83}},\ \bibinfo {pages} {064002} (\bibinfo {year}
  {2011}),\ \Eprint{http://arxiv.org/abs/1009.4932}{arXiv:1009.4932
  [astro-ph.HE]}%
  \bibAnnoteFile{NoStop}{Paschalidis:2010dh}%
\bibitem{Paschalidis:2011ez}%
  \BibitemOpen
  \bibfield{author}{%
  \bibinfo {author} {\bibfnamefont{V.}~\bibnamefont{Paschalidis}}, \bibinfo
  {author} {\bibfnamefont{Y.~T.}\ \bibnamefont{Liu}}, \bibinfo {author}
  {\bibfnamefont{Z.}~\bibnamefont{Etienne}},\ and\ \bibinfo {author}
  {\bibfnamefont{S.~L.}\ \bibnamefont{Shapiro}},\ }%
  \bibfield{journal}{%
  \Doi{10.1103/PhysRevD.84.104032}{\bibinfo {journal} {Phys.Rev.}}\ }%
  \textbf{\bibinfo {volume} {D84}},\ \bibinfo {pages} {104032} (\bibinfo {year}
  {2011}),\ \Eprint{http://arxiv.org/abs/1109.5177}{arXiv:1109.5177
  [astro-ph.HE]}%
  \bibAnnoteFile{NoStop}{Paschalidis:2011ez}%
\bibitem{EAST}%
  \BibitemOpen
  \bibfield{author}{%
  \bibinfo {author} {\bibfnamefont{W.~E.}\ \bibnamefont{East}}, \bibinfo
  {author} {\bibfnamefont{F.~M.}\ \bibnamefont{Ramazanoglu}},\ and\ \bibinfo
  {author} {\bibfnamefont{F.}~\bibnamefont{Pretorius}},\ }%
  \bibfield{journal}{%
  \Doi{10.1103/PhysRevD.86.104053}{\bibinfo {journal} {Phys.Rev.}}\ }%
  \textbf{\bibinfo {volume} {D86}},\ \bibinfo {pages} {104053} (\bibinfo {year}
  {2012}),\ \Eprint{http://arxiv.org/abs/1208.3473}{arXiv:1208.3473 [gr-qc]}%
  \bibAnnoteFile{NoStop}{EAST}%
\bibitem{TwoPunctures}%
  \BibitemOpen
  \bibfield{author}{%
  \bibinfo {author} {\bibfnamefont{M.}~\bibnamefont{Ansorg}}, \bibinfo {author}
  {\bibfnamefont{B.}~\bibnamefont{Bruegmann}},\ and\ \bibinfo {author}
  {\bibfnamefont{W.}~\bibnamefont{Tichy}},\ }%
  \bibfield{journal}{%
  \Doi{10.1103/PhysRevD.70.064011}{\bibinfo {journal} {Phys.Rev.}}\ }%
  \textbf{\bibinfo {volume} {D70}},\ \bibinfo {pages} {064011} (\bibinfo {year}
  {2004}),\ \Eprint{http://arxiv.org/abs/gr-qc/0404056}{arXiv:gr-qc/0404056
  [gr-qc]}%
  \bibAnnoteFile{NoStop}{TwoPunctures}%
\bibitem{Ajith}%
  \BibitemOpen
  \bibfield{author}{%
  \bibinfo {author} {\bibfnamefont{P.}~\bibnamefont{{Ajith}}}, \bibinfo
  {author} {\bibfnamefont{S.}~\bibnamefont{{Babak}}}, \bibinfo {author}
  {\bibfnamefont{Y.}~\bibnamefont{{Chen}}}, \bibinfo {author}
  {\bibfnamefont{M.}~\bibnamefont{{Hewitson}}}, \bibinfo {author}
  {\bibfnamefont{B.}~\bibnamefont{{Krishnan}}}, \bibinfo {author}
  {\bibfnamefont{A.~M.}\ \bibnamefont{{Sintes}}}, \bibinfo {author}
  {\bibfnamefont{J.~T.}\ \bibnamefont{{Whelan}}}, \bibinfo {author}
  {\bibfnamefont{B.}~\bibnamefont{{Br{\"u}gmann}}}, \bibinfo {author}
  {\bibfnamefont{P.}~\bibnamefont{{Diener}}}, \bibinfo {author}
  {\bibfnamefont{N.}~\bibnamefont{{Dorband}}}, \bibinfo {author}
  {\bibfnamefont{J.}~\bibnamefont{{Gonzalez}}}, \bibinfo {author}
  {\bibfnamefont{M.}~\bibnamefont{{Hannam}}}, \bibinfo {author}
  {\bibfnamefont{S.}~\bibnamefont{{Husa}}}, \bibinfo {author}
  {\bibfnamefont{D.}~\bibnamefont{{Pollney}}}, \bibinfo {author}
  {\bibfnamefont{L.}~\bibnamefont{{Rezzolla}}}, \bibinfo {author}
  {\bibfnamefont{L.}~\bibnamefont{{Santamar{\'{\i}}a}}}, \bibinfo {author}
  {\bibfnamefont{U.}~\bibnamefont{{Sperhake}}},\ and\ \bibinfo {author}
  {\bibfnamefont{J.}~\bibnamefont{{Thornburg}}},\ }%
  \bibfield{journal}{%
  \Doi{10.1103/PhysRevD.77.104017}{\bibinfo {journal} {\prd}}\ }%
  \textbf{\bibinfo {volume} {77}},\ \bibinfo {pages} {104017} (\bibinfo {month}
  {May}\ \bibinfo {year} {2008})%
  \bibAnnoteFile{NoStop}{Ajith}%
\bibitem{NINJA2}%
  \BibitemOpen
  \bibfield{author}{%
  \bibinfo {author} {\bibfnamefont{P.}~\bibnamefont{Ajith}}, \bibinfo {author}
  {\bibfnamefont{M.}~\bibnamefont{Boyle}}, \bibinfo {author}
  {\bibfnamefont{D.~A.}\ \bibnamefont{Brown}}, \bibinfo {author}
  {\bibfnamefont{B.}~\bibnamefont{Brugmann}}, \bibinfo {author}
  {\bibfnamefont{L.~T.}\ \bibnamefont{Buchman}}, \emph{et~al.},\ }%
  \bibfield{journal}{%
  \Doi{10.1088/0264-9381/29/12/124001}{\bibinfo {journal} {Class.Quant.Grav.}}\
  }%
  \textbf{\bibinfo {volume} {29}},\ \bibinfo {pages} {124001} (\bibinfo {year}
  {2012}),\ \Eprint{http://arxiv.org/abs/1201.5319}{arXiv:1201.5319 [gr-qc]}%
  \bibAnnoteFile{NoStop}{NINJA2}%
\bibitem{ET}%
  \BibitemOpen
  \bibfield{author}{%
  \bibinfo {author} {\bibfnamefont{F.}~\bibnamefont{Loffler}}, \bibinfo
  {author} {\bibfnamefont{J.}~\bibnamefont{Faber}}, \bibinfo {author}
  {\bibfnamefont{E.}~\bibnamefont{Bentivegna}}, \bibinfo {author}
  {\bibfnamefont{T.}~\bibnamefont{Bode}}, \bibinfo {author}
  {\bibfnamefont{P.}~\bibnamefont{Diener}}, \emph{et~al.},\ }%
  \bibfield{journal}{%
  \Doi{10.1088/0264-9381/29/11/115001}{\bibinfo {journal} {Class.Quant.Grav.}}\
  }%
  \textbf{\bibinfo {volume} {29}},\ \bibinfo {pages} {115001} (\bibinfo {year}
  {2012}),\ \Eprint{http://arxiv.org/abs/1111.3344}{arXiv:1111.3344 [gr-qc]}%
  \bibAnnoteFile{NoStop}{ET}%
\bibitem{Boyd}%
  \BibitemOpen
  \bibfield{author}{%
  \bibinfo {author} {\bibfnamefont{J.~P.}\ \bibnamefont{Boyd}},\ }%
  \emph{\bibinfo {title} {Chebyshev and Fourier Spectral Methods: Second
  Revised Edition}}\ (\bibinfo {publisher} {Dover Books on Mathematics},\
  \bibinfo {year} {2001})%
  \bibAnnoteFile{NoStop}{Boyd}%
\end{thebibliography}%

\end{document}